\documentclass[10pt,a4paper]{article}
\usepackage[utf8]{inputenc}
\usepackage[T1]{fontenc}
\usepackage{tabularx}
\usepackage{geometry}
\usepackage{setspace}
\usepackage{titlesec}
\usepackage{amsmath}
\usepackage{hyperref}
\usepackage{graphicx}

\titleformat{\section}{\normalfont\large\bfseries}{\thesection.}{1em}{}
\titleformat{\subsection}{\normalfont\normalsize\bfseries}{\thesubsection.}{1em}{}

\title{Physical fitness post \( \mathrm{VO_{2}max} \) - a computational framework}

\author{
J. Borresen$^{1}$ and H.L. Burger\\[1ex]
{\small
$^{1}$Sport, Exercise Medicine and Lifestyle Institute, University of Pretoria, South Africa
}
}

\date{}

\begin{document}

\maketitle

\section*{}

This paper critically examines the conceptual and methodological limitations underlying the current understanding and application of \( \mathrm{VO_{2}max} \) in exercise science and physical activity prescription. Despite the establishment of WHO guidelines on physical activity, population-level adherence remains low. A key contributing factor is the continued reliance on \( \mathrm{VO_{2}max} \) as a central measure of physical fitness - an index that represents statistical relationships within large populations that provides limited insight or utility at the individual level. The authors demonstrate that the intrinsic constraints of \( \mathrm{VO_{2}max} \) make it unsuitable for the development of precise computational models capable of informing individualized exercise prescription. The paper reviews fundamental principles linking health, fitness, and exercise intensity, identifying critical gaps in how these constructs are conceptualized and operationalized. In response, alternative theoretical and methodological frameworks are proposed to more accurately capture the complex interplay between physiological and psychological factors that determine exercise performance and adherence. Building on this conceptual foundation, the authors introduce a novel computational approach that mathematically models these complex interdependencies and the various variables that affect adherence outcomes.  A comprehensive empirical research framework and corresponding methodologies for deriving the required model equations are presented. Successful empirical validation of this model could provide a transformative step toward personalized, adaptive training systems—enhancing both the efficacy and long-term adherence of health-oriented exercise programs.

\section*{Introduction}

Little progress has been made in implementing WHO guidelines on physical activity in the general population, indicating that current approaches are ineffective.  Many existing physical fitness interventions rely on metrics such as \( \mathrm{VO_{2}max} \) that were developed in the mid-20th century.  These concepts represent statistical associations observed across large populations with little utility in the analysis and prescription of individual physical activity.  Effective solutions for 21st century challenges will be accelerated by leveraging 21st century methodologies and technologies.  The authors will review the utility of \( \mathrm{VO_{2}max} \) in modern science and demonstrate that its many limitations render it unsuitable to developing accurate computational models for effective physical activity analysis and prescription. The validity of fundamental principles underpinning the current approach to physical fitness is reviewed.  Alternative approaches and principles are proposed to support accurate 21st century computational models to analyse and prescribe physical activity for individuals.  Currently the relationships between health, fitness and physical activity-related intensity are poorly defined and the existing definitions of physical fitness and physical activity-related intensity are inadequate for advanced modelling. Underlying phenomena require definition in a manner conducive to accurate and objective quantification before it may be effectively represented as a variable in a computational model. The authors will propose more comprehensive definitions for physical fitness and physical activity-related intensity, as well as a theoretical framework for the accurate modelling, measurement and prediction of the relationships between physical activity, fitness and health. The critical interplay between physiological and psychological factors in effective exercise prescription will be examined and the complex relationships between the many interdependent variables mapped. The necessity for leveraging computational models to effectively manage the diverse interdependent variable relationships and large data point collections required for optimal physical activity prescription will be demonstrated.  Algorithms and equations are developed to prescribe physical activity on an individual intra session and inter session basis.  Empirical research methodologies are designed to accurately measure the interdependent variable relationships that constitute the required large data point collections.

\section*{General limitations in the application of \( \mathrm{VO_{2}max} \)}

The American College of Sports Medicine (ACSM) states that cardiorespiratory fitness (CRF) “reflects the functional capabilities of the heart, blood vessels, lungs, and skeletal muscles to transport and utilize oxygen to perform physical work”\textsuperscript{1} and is “related to the ability to perform large muscle, dynamic, moderate-to-vigorous intensity exercise for prolonged periods of time”.\textsuperscript{1} It further states that cardiorespiratory endurance is “the ability of the respiratory and circulatory system to supply oxygen during sustained physical activity”\textsuperscript{1} and that: “\( \mathrm{VO_{2}max} \) is accepted as the criterion measure of CRF”.\textsuperscript{1} A \( \mathrm{VO_{2}max} \) test is a maximal test and measures the ability of the respiratory and circulatory system to supply oxygen during maximal effort. A \( \mathrm{VO_{2}max} \) test does not measure the ability of the respiratory and circulatory system to supply oxygen during sustained physical activity. Therefore, if it is accepted that \( \mathrm{VO_{2}max} \) is the criterion measure of CRF, it must also be accepted that the term CRF reflects the functional capabilities of the heart, blood vessels, lungs, and skeletal muscles to transport and utilize oxygen to perform physical work in the limited case of maximal effort only. The scope of the term CRF cannot be expanded to “the ability to perform large muscle, dynamic, moderate-to-vigorous intensity exercise for prolonged periods of time”, nor can it be expanded to “the ability of the respiratory and circulatory system to supply oxygen during sustained physical activity”, unless the statement that “\( \mathrm{VO_{2}max} \) is the criterion measure of CRF” is abandoned.  In terms of these definitions, cardiorespiratory fitness cannot be equated with cardiorespiratory endurance, nor can it be considered a reflection of the ability to perform “exercise for prolonged periods of time”. It should be noted that cardiorespiratory endurance is only one of many components that limit the ability to perform moderate-to-vigorous intensity exercise for prolonged periods of time. Overall endurance performance will be limited by the component with the lowest level of endurance capacity.  Unless it is known that the cardiorespiratory component is the component with the lowest endurance capacity, which in the general case cannot be assumed, cardiorespiratory endurance will not predict overall endurance performance at the individual level.\textsuperscript{2} This is contrary to common misconception where cardiorespiratory endurance or cardiorespiratory fitness is used interchangeably with endurance performance. The ability to sustain moderate-to-vigorous intensity physical activity for prolonged periods of time is primarily a function of conditioning, not \( \mathrm{VO_{2}max} \). In the context of this paper, we define conditioning as “physiological adaptation brought about by increased demands of physical activity”. Conditioning is physical activity type specific. When we introduce the terms prolonged periods of time, we introduce the concept of endurance. In the context of this paper, we consider physical activity endurance as the ability to sustain physical activity at any intensity for prolonged periods of time. It may be accurately stated that \( \mathrm{VO_{2}max} \) is a statistical indicator of physical activity endurance for moderate-to-vigorous intensity physical activity. It is generally accepted that there is a correlation between averaged \( \mathrm{VO_{2}max} \) and averaged endurance performance\textsuperscript{3} at moderate-to-vigorous intensity for sufficiently large populations. This is to be expected, considering that any form of muscular conditioning over prolonged periods of time at moderate-to-vigorous intensity will place increased demands for oxygen on the cardiorespiratory system, which will necessarily induce conditioning of the cardiorespiratory system in most cases. Consequently, \( \mathrm{VO_{2}max} \) should be considered an indirect and limited statistical predictor of physical activity endurance in large populations, the utility of which decreases with a decrease in intensity and an increase in duration. This decrease in utility is due to the minimal demand placed on the cardiorespiratory system at low intensities, with resultant minimal adaptation of the cardiorespiratory system.  In the case of individuals, the utility of \( \mathrm{VO_{2}max} \) in predicting physical activity endurance performance decreases even further, to the point that it may be considered of no value, as demonstrated by Clarke et al.\textsuperscript{4} The authors do not agree with the conclusion by Clarke et al that distance achieved represents “submaximal cardiorespiratory fitness”, but instead are of the view that it represents endurance performance, albeit of a short duration. Nevertheless, the data clearly demonstrates that there is no correlation between \( \mathrm{VO_{2}max} \) and endurance performance in individuals. Data from a study by Vollaard et al\textsuperscript{3} supports this conclusion.

\section*{Unreliability of \( \mathrm{VO_{2}max} \) as a measure of peak cardiorespiratory capacity}

A number of additional difficulties exist for \( \mathrm{VO_{2}max} \). It cannot be considered an objective measure. Its measurement is 100\% dependent on self-reported exhaustion, which relies on subjective perception by the individual. This perception is among other things a function of the motivation and capacity of the individual to endure discomfort and pain. The ACSM reports a higher average \( \mathrm{VO_{2}max} \) when measured on a treadmill than a stationary bicycle.\textsuperscript{1} If we are to accept that the two activities are unlikely to affect the functioning of the cardiorespiratory system in a significantly dissimilar manner (other than possible minor inhibition of the ventilatory system\textsuperscript{5}), this discrepancy can only be accounted for by the fact that the \( \mathrm{VO_{2}max} \) measurement is limited by the demand for O2 and not the ability of the cardiorespiratory system to supply it.  A review by Millet et al\textsuperscript{5} reported that in triathletes with highly conditioned demand systems for both running and cycling, equivalent \( \mathrm{VO_{2}max} \) readings can be obtained for both activities. At the very least, we must conclude that for an average individual, a \( \mathrm{VO_{2}max} \) test performed on a stationary bicycle is not a measure of cardiorespiratory capacity at all, but rather the demand capacity of other physiological components that lead to exhaustion. Additionally, it raises the question whether a \( \mathrm{VO_{2}max} \) test performed on a treadmill is a true reflection of cardiorespiratory capacity in most cases, even assuming that exhaustion is not reported prematurely due to psychological factors. In order to increase the probability that a \( \mathrm{VO_{2}max} \) test will reflect the peak capacity of the cardiorespiratory system, the type of activity chosen must be in accordance with the activity that the physiology of the subject has been most highly conditioned to.

\section*{Specific errors in the measurement of \( \mathrm{VO_{2}max} \) due to body composition}

In addition to the unsuitability of \( \mathrm{VO_{2}max} \) as a measure of cardiorespiratory endurance as defined by the ACSM, it is also deeply flawed as a measure of “the ability to supply oxygen to the musculoskeletal system to meet maximum demand during maximum effort”. This is a consequence of the units of measure of CRF, namely ml/kg/min where kg is the body weight of the individual. This can be demonstrated in the following manner: Assume an individual with a lean body mass of 70kg, a body fat mass of 30kg and a measured \( \mathrm{VO_{2}max} \) of 30ml/kg/min. Now assume a theoretical scenario where this individual loses 20kg of body fat without any physiological changes in the cardiorespiratory system (e.g. liposuction). A \( \mathrm{VO_{2}max} \) test post weight loss will reflect a measurement of 37.5ml/kg/min, an increase in \( \mathrm{VO_{2}max} \) of 25\% purely as a consequence of undergoing liposuction. Such a result is clearly implausible. It is self-evident that when normalising \( \mathrm{VO_{2}max} \) to an individual, lean body mass only should be used in the calculation.  As a general principle, the authors are of the view that when normalizing absolute functional measurements in a population to the size of an individual in that population, the measurement should be normalized to the size of the functional physiology that produces the measured phenomenon.

\section*{Specific errors in the measurement of \( \mathrm{VO_{2}max} \) due to body composition extended to METs}

A similar difficulty arises when considering METs, which is also normalized to body weight. The ACSM presents METs as a measure of intensity\textsuperscript{1} and the Compendium of Physical Activities referenced by the ACSM\textsuperscript{6} (updated in 2024\textsuperscript{7}), refers to a value of 6 METs when cycling at 90-100 Watts on a stationary bicycle. Let us now consider the case of an individual with a lean body mass of 70kg cycling at 100W mechanical power on a stationary bicycle and consuming 6 METs of power. 1 MET equates to approximately 1.162 W/kg.  The total power consumption of the individual is 6 × 1.162 W/kg × 70kg = 488.04W. Assume the individual gains 30kg body fat before again cycling at 100W mechanical power. For purposes of illustration we assume the weight gain by the individual will not significantly affect exercise efficiency on a stationary bicycle. When calculating the exercise intensity of the individual post weight gain in METs, we find an intensity of 488.04W / 100kg / 1.162 = 4.2 METs. The current definition of exercise intensity in METs therefore predicts that an individual will exercise at a 30\% lower intensity post weight gain, when performing the same activity and mechanical work, purely by virtue of the fact that the individual has gained 30kg of body fat. Such a prediction is absurd and again confirms that the current practice of calculating exercise intensity from total body weight is deeply flawed. When normalising METs to an individual, lean body mass only should be used in the calculation.

\section*{Specific errors in exercise prescription in METs due to body composition}

The error in prescribing exercise in METs based on total body weight is further compounded when considering the type of exercise performed. In the previous example exercise on a stationary bicycle was deliberately chosen to control for the effect of body fat on the work rate associated with the exercise activity in question. Let us now turn to the prescription of vigorous exercise in the form of walking briskly uphill. For purely illustrative purposes and ignoring possible changes in exercise efficiency, we will assume (based on the laws of physics) that the power required to move a body mass up a constant grade at a fixed velocity increases proportionally with the mass of the body. Let us consider the case of an individual A with a lean and total body mass of 70kg ascending a moderate grade at a moderate pace and an energy consumption of 7METs (“Walking - climbing hills, no load, 6 to 10\% grade, moderate-to-brisk pace”\textsuperscript{7}). Let us further consider the case where individual A gains 20kg body fat before again ascending the same moderate grade at the same moderate pace. From our previous assumptions, we conclude theoretically that the energy consumption during the second ascent will rise to 7METs × (90kg/70kg) = approximately 9METs. From the aforementioned examples we see that the true intensity (normalised to lean body mass) of the prescribed physical activity in this case increases by 28.5\% for an individual that has gained 20kg of fat mass over and above a lean body mass of 70kg. There is no sound reason why an individual with a higher body fat percentage should be prescribed exercise at an intensity higher than that prescribed to a lean person. Medical risk factors in obese individuals indicate the contrary. The validity of this example is evident in the Compendium\textsuperscript{7} itself, which ascribes higher MET values to walking activities carrying loads, demonstrating that an increase in weight other than lean body mass will increase the intensity (energy cost/kg of lean body mass) of the activity. It is clear that many physical activities cannot be prescribed to individuals in the general population based on a theoretical MET value that disregards body composition. Exercise prescription must be adjusted appropriately for a range of body compositions for each type of activity, where applicable.

\section*{A brief perspective on the consequences of inappropriate application of \( \mathrm{VO_{2}max} \)}

The consequences of relying on total body mass to measure \( \mathrm{VO_{2}max} \) are far reaching.  A great many studies have investigated the correlation between health outcomes and \( \mathrm{VO_{2}max} \). Since the ACSM defines \( \mathrm{VO_{2}max} \) as a measurement of CRF, those health outcomes have been associated with CRF. As demonstrated in this paper, such an interpretation is often not appropriate.  Individuals in global urban populations contain on average approximately 40\% body fat (females) and approximately 28\% body fat (males) (US DHHS 1999-2004 data\textsuperscript{8}). The inclusion of body fat in the units of measure for \( \mathrm{VO_{2}max} \) has the unintended consequence of misinterpreting a correlation between body fat mass and health as a correlation between CRF and health.

The ACSM relies on studies that compare the correlation between mortality and \( \mathrm{VO_{2}max} \) (cardiorespiratory fitness); and mortality and physical activity volume respectively.  It has been found that the difference between the cardiovascular disease mortality risk in the lowest percentile versus the top percentile of \( \mathrm{VO_{2}max} \) is greater than the difference between the mortality risk in the bottom percentile versus the top percentile of physical activity volume.\textsuperscript{9} If these conclusions were to be accepted, it is apparent that a maximal \( \mathrm{VO_{2}max} \) test is a better predictor for mortality and consequently health outcomes than an assessment of physical activity volume. However, the question arises what the utility of these two respective measures is in the context of physical activity. If the objective is to prescribe physical activity and monitor the effectiveness thereof in reducing mortality, it needs to be considered that \( \mathrm{VO_{2}max} \) does not accurately reflect the relationship between physical activity and health outcomes. This is a consequence of at least two effects:  In the case of utilizing lean body mass to measure \( \mathrm{VO_{2}max} \), the lowest percentile will include relatively more subjects with underlying adverse cardiorespiratory conditions. While diagnosed conditions are in most cases screened out in studies, this is not the case for undiagnosed conditions. These conditions will contribute to greater mortality in the bottom percentile. While this effect results in \( \mathrm{VO_{2}max} \) being a superior predictor of mortality than physical activity volume, it is inferior in predicting the relationship between physical activity and mortality. In the case of utilizing total body mass to measure \( \mathrm{VO_{2}max} \) (as is usual), the bottom \( \mathrm{VO_{2}max} \) percentile will include relatively more obese subjects due to the effect of body fat on the standard \( \mathrm{VO_{2}max} \) measure (as previously demonstrated in this paper). It is well established that increased obesity results in increased mortality.\textsuperscript{10} While it is again the case that this second effect results in \( \mathrm{VO_{2}max} \) being a superior predictor of mortality than physical activity volume, it exacerbates the inferiority of \( \mathrm{VO_{2}max} \) in predicting the relationship between physical activity and health outcomes. We consequently conclude that for the purpose of prescribing physical activity and monitoring the effectiveness thereof in reducing mortality, assessments of physical activity volume would be superior to \( \mathrm{VO_{2}max} \).

\section*{Improved measures of physical activity-related intensity developed from first principles}

In order to effectively prescribe physical activity, a model is required that accurately maps the relationships between the various relevant physiological phenomena.  Considering the presented limitations of \( \mathrm{VO_{2}max} \), it is apparent that such a model cannot rely on \( \mathrm{VO_{2}max} \) related measurements, necessitating the development of new and improved definitions and measurements. In table 5.2 of the ACSM’s Guidelines for Exercise Testing and Prescription\textsuperscript{1} the ACSM presents essentially three types of intensity, perceived intensity (RPE), relative intensity (\%\( \mathrm{VO_{2}} \)R, \%HRR) and absolute intensity (METs). A rudimentary attempt is made to demonstrate how these three intensities may be estimated from each other, but no attempt is made to clearly define any of them, leaving open the question:  What is physical activity-related intensity really? This question is particularly relevant as the three “intensities” are often used interchangeably. This lack of clear definition of what is being measured in absolute terms is extended to research on the relationship between physical activity-related intensity and other phenomena. The authors will demonstrate that it is critical that physical activity-related intensity is clearly defined in absolute terms when studying its relationships with phenomena such as health, physiological adaptation, performance and motivation. Consequently, the authors propose new and more precise definitions for physical activity-related intensity in order to meaningfully develop their research on physical activity prescription and measurement. Before we define the intensity of physical activity, it is necessary to adequately define physical activity itself. The ACSM defines physical activity as follows: “Physical activity is defined as any bodily movement produced by the contraction of skeletal muscles that results in an increase in caloric requirements over resting energy expenditure”.\textsuperscript{1} The only absolute measure for the intensity of physical activity in SI units is power. The current ACSM definition of physical activity requires physical activity-related intensity to be expressed in terms of resting energy expenditure (REE) as well as the absolute measured physical activity-related intensity. Therefore, REE must be known.  For an individual, REE must be either estimated or measured. Estimation of REE in an individual is not accurate, resulting in inaccuracies of the determined physical activity-related intensity. Measurement of REE is a time consuming and complex process (especially in the more accurate case of basal metabolic rate), rendering it impractical in most use cases where physical activity-related intensity must be determined. In order to overcome these limitations in the measurement of physical activity in the context of its current ACSM definition, the authors propose that physical activity be defined more precisely and practically in terms of energy as follows: 

\begin{enumerate}
    \item\textit{“Physical activity (PA) is the process of transforming chemical energy to kinetic energy by the physiology of animalia”.}
\end{enumerate}

This definition diverges from the ACSM definition in that for example the process of breathing is considered physical activity regardless of rate, whereas the ACSM definition does not consider the process of breathing at a low rate (at REE) to be physical activity, but at a higher rate (above REE) it is considered physical activity. The authors are of the view that physical activity is more appropriately defined by its nature than by its rate.

\begin{enumerate}
    \setcounter{enumi}{1}
    \item\textit{“Consumption energy (CE) is the total chemical energy consumed in order to produce kinetic energy”.}
    \item\textit{“Physical activity is quantified by expressing it as consumption energy”.}
    \item\textit{“Physical activity intensity (PAI) can be expressed as consumption intensity (CI), which is the rate at which physical activity consumes chemical energy, expressed as power/kg of lean body mass”.}
    \item\textit{“Physical activity kinetic energy is the kinetic energy produced by physical activity”.}
    \item\textit{“Kinetic energy intensity (KI) is the rate at which physical activity produces kinetic energy, expressed as power/kg of lean body mass”.}
\end{enumerate}

The process of converting chemical energy to kinetic energy is not efficient. Thus the kinetic energy produced does not equal the total chemical energy consumed.  

\begin{enumerate}
    \setcounter{enumi}{6}
    \item\textit{“Physical activity efficiency (Epa) is the ratio of kinetic energy produced to total chemical energy consumed”.}
    \item\textit{“Performance energy (PE) is the minimum kinetic energy transferred to the external environment and the physiology of animalia in order to effect the purpose of the physical activity (task)”.}
    \item\textit{“Performance intensity (PI) is the rate of transfer of performance energy expressed as power/kg of lean body mass”.}
    \item\textit{“Performance energy efficiency (Epf) is the ratio of performance energy to kinetic energy”.}
\end{enumerate}

Performance energy efficiency is a function of the biomechanical efficiency of the type of physical activity performed, the genetic unconditioned biomechanical efficiency of the individual and the conditioned biomechanical efficiency of the individual. Physical activity efficiency is a function of the environment, the health of the individual, the inherent genetic non-biomechanical characteristics of the individual and potentially the non-biomechanical conditioning of the individual.\textsuperscript{11} The relationship between chemical energy consumption and performance energy transference is a function of physical activity efficiency and performance energy efficiency, both of which are dependent on a large number of variables. Unless all these variables are known, which cannot currently be practically achieved to a high degree of accuracy, performance energy cannot be calculated from consumption energy, nor can consumption energy be calculated from performance energy.  However, with accurate modelling and testing within a controlled environment, it may be possible to estimate one from the other.  The relationships between the various energy stages are demonstrated in Figure~1.

\begin{figure}[h!]
        \centering
        \includegraphics[width=0.55\textwidth]{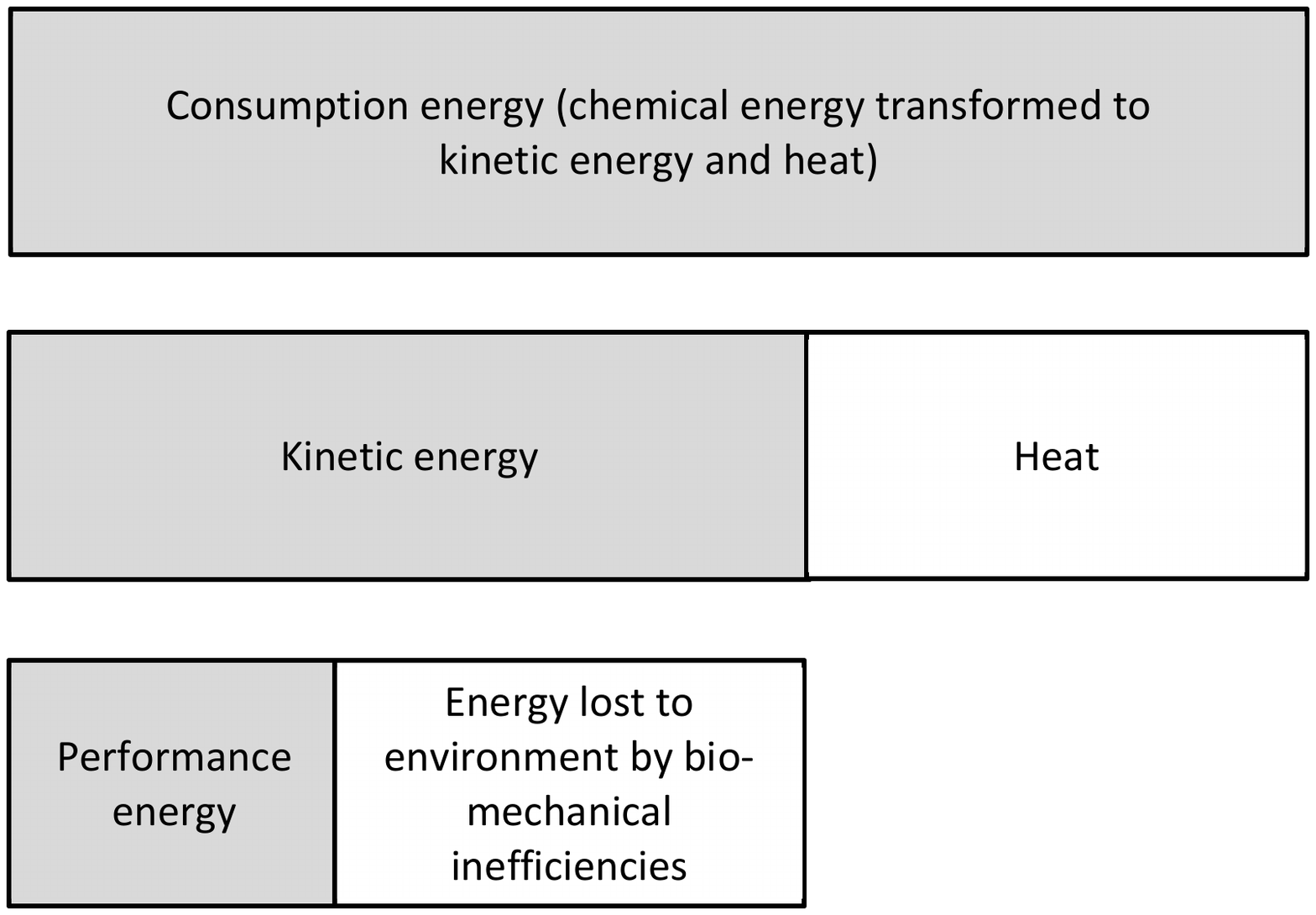}
        \caption{Transformation of energy during a specific physical activity.}
        \label{Figure 1}
    \end{figure}

In addition to the preceding physical activity-related intensities defined in terms of energy, the following physiological intensities are defined: 

\begin{enumerate}
    \setcounter{enumi}{10}
    \item\textit{“Objective experienced intensity (OEI) is the objective degree of stress on the entire physiological system caused by the performance intensity”.}
    \item\textit{“Subjective experienced intensity (RPE) is the subjective tolerance of an individual for objective experienced intensity”.}
    \item\textit{“Physiological Endurance Capacity (PEC) is the ability of the physiology to sustain task specific physical activity at a specific objective experienced intensity for a set duration”.}
    \item\textit{“Physiological Endurance Capacity (PEC) is expressed as Consumption Energy in ml/kg consumed over a Session Duration (SD) at a constant Objective Experienced Intensity (OEI).  Session duration is expressed in minutes. Objective experienced intensity is expressed as dimensionless Rated Perceived Exertion adjusted (RPEa) for Pain Tolerance (PT): PEC(SD=SDo min, RPEa =RPEao) = PECo ml/kg”.}
\end{enumerate}

Oxygen consumption (\( \mathrm{VO_{2}} \)) can serve as an accurate measure of consumption intensity, but in itself can neither predict experienced intensity nor performance intensity accurately. Figure~2 illustrates the relationships between energy and physiological intensities.

\begin{figure}[h!]
        \centering
        \includegraphics[width=0.75\textwidth]{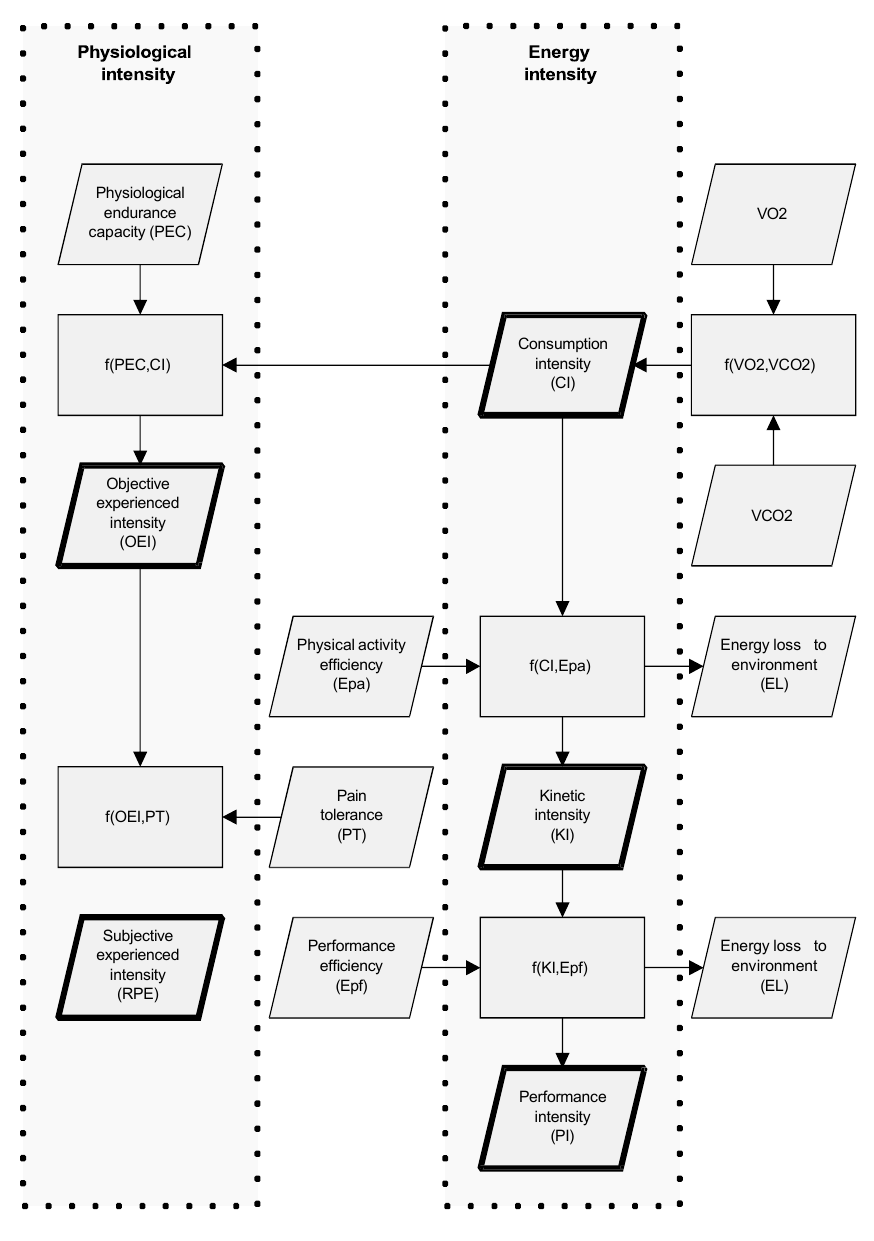}
        \caption{Physical activity intensity relationships.}
        \label{Figure 2}
    \end{figure}

\section*{The relationship between physical activity and physiological adaptation, health benefits, motivation and fitness}

It becomes apparent from the preceding section that the relationship between related phenomena and physical activity-related intensity can only be meaningfully measured if the correct definition of intensity is applied.  Upon consideration we find that: 
 
\begin{enumerate}
    \item\textit{An accurate relationship between physiological adaptation and physical activity can only be established by measuring objective experienced intensity}
    \item\textit{An accurate relationship between health benefits and physical activity can only be established by measuring consumption intensity}
    \item\textit{An accurate relationship between motivation and physical activity can only be established by measuring subjective experienced intensity}
    \item\textit{An accurate relationship between fitness and physical activity can be established by measuring performance intensity, consumption intensity or subjective experienced intensity.  Which of these measures is utilized depends on the type of fitness.}
\end{enumerate}

The above conclusions are supported as follows:\vspace{1em}

\textit{1. Physiological adaptation and objective experienced intensity}
\\Physiological adaptation is brought about by stress on the physiology.  There are various forms of physiological adaptation.  For the purposes of this paper the following definition of physiological adaptation is adopted: “Physiological adaptation is a change in endurance capacity due to exposure to a specific objective experienced intensity over a defined duration”.  An individual with greater conditioning and consequently physiological capacity will experience lower physiological stress than the same individual with less conditioning and physiological capacity for a given consumption intensity. An individual with greater biomechanical efficiency will experience lower physiological stress than the same individual with less biomechanical efficiency for a given performance intensity. The use of consumption intensity or performance intensity to measure the relationship between physical activity and physiological adaptation will result in errors due to the confounding effects of conditioning and biomechanical efficiency. Two individuals subjected to the same objective experienced intensity can exhibit substantial differences in subjective experienced intensity due to variations in tolerance for pain and discomfort between individuals.  The use of subjective experienced intensity to measure the relationship between physical activity and physiological adaptation will result in errors due to the confounding effects of pain and discomfort tolerance. In order to eliminate these and other confounders, an accurate relationship between physiological adaptation and physical activity can only be obtained if objective experienced intensity is measured.  Direct measurement of objective experienced intensity is currently not possible. This research framework will propose methodologies for investigating estimation of objective experienced intensity from subjective experienced intensity (RPE), heart rate, blood lactate and pain tolerance questionnaires.\\
  
\textit{2. Health benefits and consumption intensity}
\\The physiology of an organism will evolve to achieve the optimal condition required to meet the demands of its evolutionary environment. An environment that diverges from the evolutionary environment will result in a sub-optimal physiological condition of the organism.  This will manifest as adverse health outcomes. The relationship between health and physical activity is evolutionary in nature. Human physiology has been optimized through evolution to perform a range of physical activities in order for humans to survive under primitive conditions. Modern technology has obviated the need for many of those activities. The relative recentness of technology in terms of evolutionary time scales has resulted in a divergence between the evolutionary environment and the environment of a modern urban human. This divergence in physical activity requirements and consequently physical activity, manifests as adverse health outcomes in inactive humans. The degree to which all the physiological components are active will determine health outcomes. This includes pre-kinetic physiological components responsible for consuming chemical energy prior to transformation to kinetic energy. In order to include these pre-kinetic components when assessing activity, it is necessary to measure consumption intensity. Neither objective experienced intensity, subjective experienced intensity nor performance intensity can be used to accurately measure the activity of both pre- and post- kinetic physiological components. Given the relationship between O2 consumption and consumption intensity is well established in aerobic physical activity, an increase in \( \mathrm{VO_{2}} \) (ml/kg/min) represents an accurate measure of an increase in consumption intensity for a corresponding increase in steady state aerobic physical activity. To further demonstrate the aforementioned principles, we present examples of errors that occur when the incorrect physical activity-related intensity is measured:\\

\textit{2.1 Health benefits and subjective experienced intensity}
\\Consider the case where the relationship between physical activity and health benefits for two individuals are studied, one highly conditioned and the other highly unconditioned. Both individuals report 60 min of exercise per week at an RPE of 14. However, the conditioned individual performs physical activity equivalent to running 10km in 60min, while the unconditioned individual performs physical activity equivalent to walking 2km in 60min. It is evident that the conditioned individual will achieve superior health benefits to the unconditioned individual at the same RPE, duration and frequency. It is common practice to use RPE-based questionnaires (i.e. low, moderate or vigorous) to collect data on the relationship between physical activity-related intensity and health benefits. Consequently, depending on the conditioning of the study population, widely divergent relationships between health benefits and physical activity-related intensity will be observed if physical activity-related intensity is measured using subjective experienced intensity (RPE). This is a consequence of the confounding influence of the conditioning of the study population which obscures the true relationship between physical activity and health benefits. If, however, consumption intensity is used to measure the relationship between physical activity-related intensity and health benefits in our example case, we will note that for the highly conditioned individual high values for consumption intensity and health benefits will be observed, while for the unconditioned individual low values of consumption intensity and health benefits will be observed, preserving a consistent relationship model between physical activity-related intensity and health benefits regardless of conditioning.\\

\textit{2.2 Health benefits and objective experienced intensity}
\\The limitations demonstrated in using subjective experienced intensity to measure the relationship between physical activity and health benefits extends to objective experienced intensity. Objective experienced intensity eliminates the confounding effect of differences in discomfort and pain tolerance between individuals and is therefore a more accurate measure of the relationship between physical activity and health than subjective experienced intensity. However, the limitations demonstrated in paragraph 2.1 in the case of conditioned and unconditioned populations apply equally when objective experienced intensity is used to measure the relationship between physical activity and health benefits. Analogously, when consumption intensity is measured to determine the relationship between physical activity and health benefits, these confounding effects are eliminated.\\

\textit{2.3 Health benefits and performance intensity}
\\Consider two individuals, one a highly skilled former elite swimmer and the other with no swim training and poor skill. Over a period of ten years the only exercise performed by them is the swimming of 750m in 30 minutes twice a week, resulting in identical performance. This will equate to 60 minutes a week of low intensity exercise for the skilled individual and 60 minutes a week of vigorous exercise for the unskilled individual due to the differences in their biomechanical efficiencies. The unskilled individual will achieve significantly superior health benefits to the skilled individual despite both exercising at the same performance intensity. If, however, we used consumption intensity to measure the relationship between physical activity-related intensity and health benefits, we will note that for the highly skilled individual low values for consumption intensity and health benefits will be observed, while for the unskilled individual high values of consumption intensity and health benefits will be observed, preserving a consistent relationship model between physical activity-related intensity and health benefits regardless of skill.\\
  
\textit{3. Motivation and subjective experienced intensity}
\\It is well accepted that an increase in pain and discomfort will decrease the motivation of an individual to sustain physical activity. In order to accurately establish the relationship between motivation and physical activity-related intensity in an individual, the tolerance of that individual for objective experienced intensity should be taken into account. Accordingly, physical activity should be measured in terms of subjective experienced intensity and not objective experienced intensity. Neither consumption intensity nor performance intensity can be used to establish an accurate relationship between motivation and physical activity-related intensity. For example, an individual with greater conditioning will experience lower levels of discomfort than the same individual with less conditioning, for a given consumption intensity and pain tolerance. An individual with greater bio-mechanical efficiency will experience lower levels of discomfort than the same individual with less bio-mechanical efficiency, for a given performance intensity and pain tolerance. The use of consumption intensity or performance intensity to measure the relationship between physical activity-related intensity and motivation will result in errors due to the confounding effects of conditioning, bio-mechanical efficiency and pain tolerance. A consistent relationship model between physical activity-related intensity and motivation can only be obtained by measuring subjective experienced intensity.  RPE is an example of an appropriate measure of subjective experienced intensity.\\
  
\textit{4. Fitness, performance, consumption intensity and subjective experienced intensity}
\\Before the authors can proceed to illustrate how fitness can be measured by performance intensity, consumption intensity and subjective experienced intensity, it is necessary to adequately define fitness and performance.  For the purposes of this paper, we define performance as “Physical activity executed in an arbitrarily defined manner to an objectively measured degree”.

\subsection*{Improved definition of physical fitness developed from first principles}

The ACSM describes physical fitness as follows: “Physical fitness although defined in several ways has generally been described as a set of attributes or characteristics individuals have or achieve that relate to their ability to perform physical activity and activities of daily living”.1 The vagueness of this definition renders it inadequate for scientific enquiry. In order to develop their research on physical activity-related intensity prescription and measurement, the authors are compelled to propose a more precise definition of physical fitness.\\
 
The Oxford Dictionary defines fitness as a general concept as follows:
\\“fit (adjective)
\\\indent 1. of a suitable quality, standard, or type to meet the required purpose.”\\  
It further defines purpose as follows:
\\“purpose
\\\indent 1. the reason for which something is done or created or for which something exists.”\\
 
From a general perspective, it is clear that the word fit has no meaning without the context of its required purpose. It is evident that in the physical activity field the concept of fit is only meaningful if the purpose and nature of the physical activity is clearly defined. For example an ultramarathon athlete would consider themselves unfit if they are capable of running 20km, but not 50km. Yet for health purposes, a member of the general public capable of running 20km would be considered extremely fit. The principle also applies to the various components of physical fitness, including cardiorespiratory fitness. In many types of physical activity CRF plays a subordinate role in the ability of an individual to perform that activity, in which case we should question its relevance to those activities. When we consider “fit” in the context of physical activity, the required purpose becomes not only the required type of physical activity, but the reason why it is performed. For example running to be competitive in high performance sport and running for leisure have different objectives. Recreational runners run for leisure because they enjoy it and find it pleasurable. Professional athletes run because it is a full-time occupation and they are forced to train at maximal capacity in order to be competitive. In the case of the former, running at an excessively high RPE is not enjoyable and therefore undesirable, while in the case of the latter it is essential. A leisure runner would consider themselves fit if they can run at their desired RPE for their desired distance and duration. A high-performance professional runner would consider themselves fit if they have a reasonable probability of winning a competition. It is apparent that physical activity fitness, in the true sense of the word, is not just a function of the type of physical activity performed, but also of the motivation behind the performance of the physical activity. For this reason, the authors propose four broad categories of fitness (Table 1):
\begin{table}[h!]
\centering
\begin{tabularx}{\textwidth}{|l|X|}
\hline
\textbf{Fitness Category} & \textbf{Fitness Measure} \\ \hline
High performance physical activity & Performance intensity only (maximal capacity) \\ \hline
Occupational physical activity & Performance intensity only (prescribed by employer) \\ \hline
Leisure physical activity & Performance intensity and subjective experienced intensity (both self-defined) \\ \hline
Health physical activity & Consumption intensity and subjective experienced intensity \\ \hline
\end{tabularx}
\caption{Categories and measures of fitness}
\end{table}

\noindent\textit{Measurement of high performance fitness and occupational fitness}\\
Both high performance and occupational physical activity fitness are performance based only and the required level of performance is clearly defined. Thus fitness can be accurately expressed in these cases by measured performance as a percentage of required performance.\\  

\noindent\textit{Measurement of leisure fitness}\\
Leisure physical activity fitness is a function of both performance and subjective experienced intensity (RPE). Both are defined by the individual in question based on personal preference. Leisure fitness can be accurately expressed by either exercising at a desired RPE and expressing measured performance as a percentage of desired performance; or exercising at a desired performance and expressing measured RPE as a percentage of desired RPE.\\
 
\noindent\textit{Measurement of health fitness}\\
Many research papers upon which the ACSM bases prescription of physical activity for health, do not take into account the duration (in years) for which the reported physical activity was sustained in the sample population,12 limiting the accuracy of findings on the relationship between physical activity and health. These errors are compounded in cases where the correct type of physical activity-related intensity was not considered. Significant health benefits only accrue if physical activity is sustained at the required consumption intensity, duration and frequency in the long term. The authors contend that adherence to health physical activity prescription is largely a function of motivation, which in turn is heavily influenced by subjective experienced intensity. As such the authors propose that “Health fitness is the ability of an individual to perform physical activity at the required consumption intensity, duration and frequency, at a subjective experienced intensity conducive to sustainable adherence”. In terms of this definition, it follows that health fitness should be measured in terms of both consumption intensity and subjective experienced intensity.

\subsection*{Physical activity prescription for health}
It was previously discussed that physical activity health benefits are directly associated with consumption intensity, not objective experienced intensity, subjective experienced intensity or performance intensity. Accurate prescription of physical activity for health can only be based on an accurate relationship between health benefits and physical activity-related intensity. The current practice of prescribing physical activity solely based on relative intensity (\%HRR, \%\( \mathrm{VO_{2}max} \) and RPE) is sub-optimal.  For example, a conditioned individual running 5km in 30 minutes twice a week will achieve improved health outcomes relative to an unconditioned individual walking 2km in 30 minutes twice a week.13 Both could be exercising at the same intensity expressed as a relative measure (\%\( \mathrm{VO_{2}} \)R, \%HRR, RPE). However, the unconditioned individual will perform at a lower consumption intensity than the conditioned individual, resulting in the unconditioned individual achieving reduced health outcomes relative to the conditioned individual, despite both adhering to the same relative intensity prescription. The lack of access to and the nature of \( \mathrm{VO_{2}} \) analyzers pose challenges to practical consumption intensity-based prescription of physical activity to the general population. The ACSM prescribes physical activity based on performance intensity by suggesting a relationship between consumption intensity and performance intensity through the use of the Compendium.7 However, in its current form, the Compendium fails to take into account body composition, conditioning and bio-mechanical efficiency. All of which present as significant confounding factors in the relationship between consumption intensity and performance intensity, thus rendering Compendium-based prescription inaccurate; and in some cases highly inaccurate, as previously discussed. These limitations can be overcome if an exercise prescription model is developed that enables maximal exercise intensity without resulting in session or training intervention abortion. This paper will propose a theoretical research framework and computational model for developing such a prescription model.

\subsection*{Physical activity prescription modeling for adherence}
Physical activity adherence, both short and long term, is a function of both physiological and psychological factors.  In order to accurately predict adherence computationally, these physiological and psychological factors must be accommodated in a single model. Decision making is modeled utilizing a simple cost - benefit model where perceived benefits exceeding perceived costs will result in pursuit of a specific task and perceived costs exceeding perceived benefits will result in avoidance of a specific task.  Adherence is a function of both inter-session and intra-session decision making.  Both session initiation and session completion are required for successful fitness training intervention completion and sustainability.  The relationships and factors that affect inter-session and intra-session decision making are slightly different, requiring two different but integrated models.  Self-efficacy is modeled as perceived probability of attaining the perceived benefit.  Perceived costs and benefits can be accurately measured through questionnaires.

\section*{Intra-session decision making model}

Figure 3 illustrates a computational model of intra-session decision-making. Session abortion occurs if Perceived Cost (PCo) exceeds Perceived Benefit (PB).

\begin{figure}[p]
        \centering
        \includegraphics[width=1\textwidth]{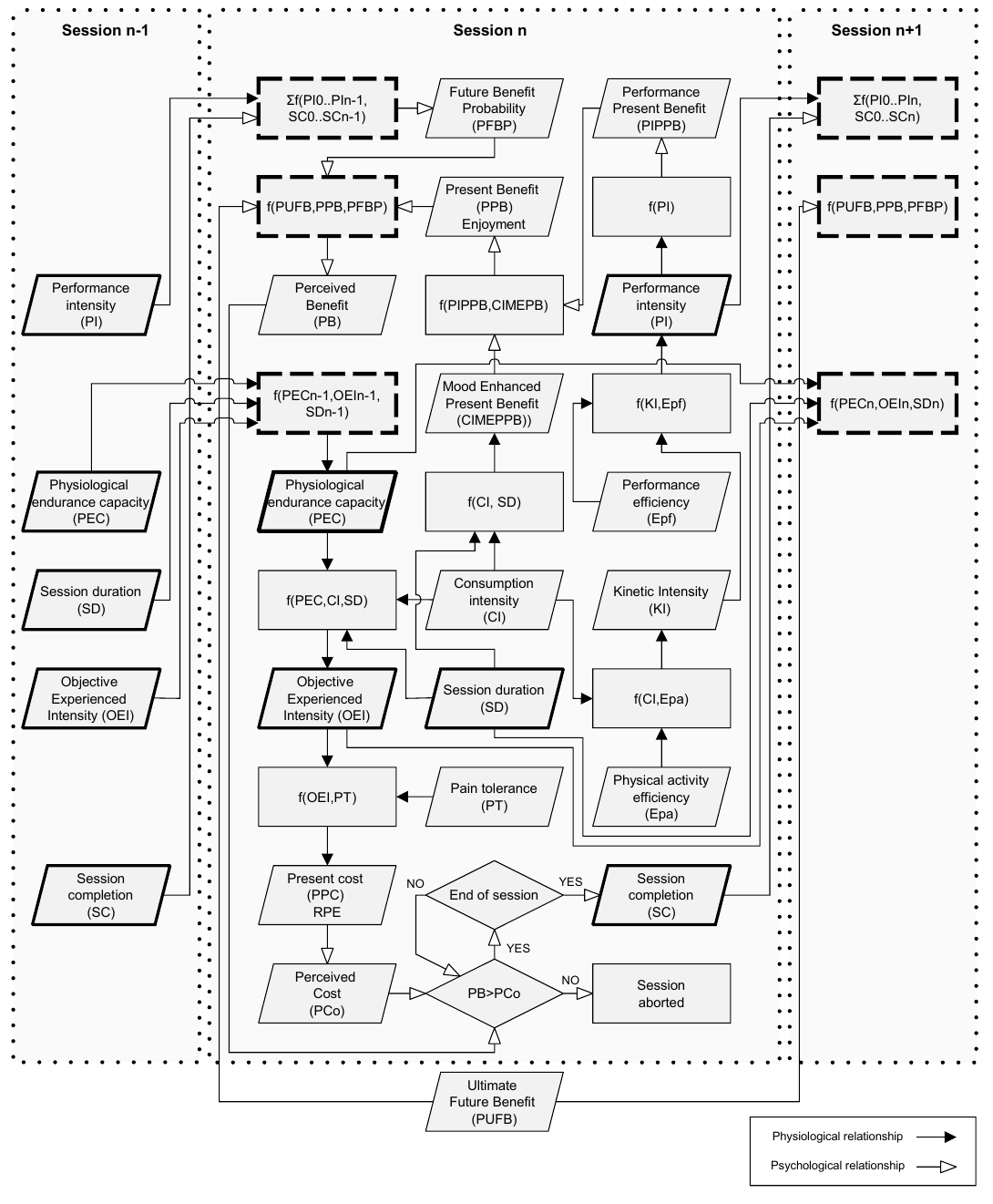}
        \caption{Intra-session decision-making model.}
        \label{Figure 3}
    \end{figure}
\vspace{1em}

Intra-session Perceived Cost (PCo) is equated to the concept of Rated Perceived Exertion (RPE), modeled as a function of Objective Experienced Intensity (OEI) and Pain Tolerance (PT). Perceived cost decreases with an increase in pain tolerance.

\[
PCo = RPE = f(OEI, PT)
\]

Intra-session Objective Experienced Intensity is modeled as a function of Physiological Endurance Capacity (PEC), Consumption Intensity (CI) and Session Duration (SD). Session Duration (SD) represents the time that has elapsed since the commencement of the exercise session. Objective Experienced Intensity increases with an increase in consumption intensity and session duration.

\[
OEI = f(PEC, CI, SD)
\]

It is well established that an unconditioned individual will exhibit greater physiological adaptation for a given OEI and duration than a conditioned individual. Consequently Physiological Endurance Capacity (PEC) is modeled as a function of prior session initial Physiological Endurance Capacity (PEC$_{n-1}$), prior session Objective Experienced Intensity (OEI$_{n-1}$) and prior session Session Duration (SD$_{n-1}$). It is calculated recursively for each session as illustrated in Figure 3. Physiological endurance capacity increases with an increase in objective experienced intensity and session duration.

\[
PEC_n = f(PEC_{n-1}, OEI_{n-1}, SD_{n-1}) = \Sigma f(PEC_0..PEC_{n-1}, OEI_0..OEI_{n-1}, SD_0..SD_{n-1})
\]

Perceived Benefit (PB) is modeled as a function of Perceived Ultimate Future Benefit (PUFB), Perceived Present Benefit (PPB) and Perceived Future Benefit Probability (PFBP). Perceived ultimate future benefit represents the perceived ultimate benefit from completing all the sessions. Perceived future benefit probability represents the perceived probability that expected future benefits will be realised and is equated to the concept of self-efficacy. Perceived present benefit represents the enjoyment derived from performance and the release of mood enhancement chemicals during exercise.\textsuperscript{14--17} Perceived benefit increases with an increase in perceived ultimate future benefit, perceived present benefit and perceived future benefit probability.

\[
PB = f(PUFB, PPB, PFBP)
\]

Perceived Present Benefit (PPB) is modeled as a function of Performance Induced Perceived Present Benefit (PIPPB) and Chemically Induced Mood Enhancement Perceived Present Benefit (CIMEPPB). Perceived present benefit increases with an increase in performance-induced perceived present benefit and chemically induced mood enhancement perceived present benefit.

\[
PPB = f(PIPPB, CIMEPPB)
\]

Performance Induced Perceived Present Benefit (PIPPB) is modeled as a function of Performance Intensity. Performance-induced perceived present benefit increases with an increase in performance intensity.\textsuperscript{18--21}

\[
PIPPB = f(PI)
\]

Performance Intensity (PI) is modeled as a function of Consumption Intensity (CI), physical activity efficiency (Epa) and performance energy efficiency (Epf). Performance intensity increases with an increase in consumption intensity, physical activity efficiency and performance energy efficiency.

\[
PI = f(CI, Epa, Epf)
\]

Chemically Induced Mood Enhancement Perceived Present Benefit (CIMEPPB) is modeled as a function of Consumption Intensity (CI) and Session Duration (SD). It increases with an increase in performance intensity and session duration.

\[
CIMEPPB = f(CI, SD)
\]

Perceived Future Benefit Probability (PFBP) is modeled as a progressively weighted function of all prior session Performance Intensities (PI) and all prior Session Completions (SC). Perceived future benefit probability increases with an increase in performance intensity and session completion.

\[
PFBP = \Sigma f(n, PI_0..PI_{n-1}, SC_0..SC_{n-1})
\]

Session Completion (SC) is modeled as a function of Perceived Cost (PCo), Perceived Benefit (PB) and Prescribed Session Duration (PSD) where prescribed session duration represents the total prescribed duration of an exercise session. Session Completion is true if PB $>$ PCo throughout SD = PSD (where PSD denotes the prescribed session duration).

\[
SC = f(PB, PCo, SD, PSD)
\]

The relationships among the above variables are formalized as follows for intra-session adherence decision-making: (Variables exerting a positive influence on the functional outcomes are denoted with a + symbol, whereas those exerting a negative influence are denoted with a - symbol. Dependent variables are presented in italicized form)

\vspace{1em}

Session completion will be achieved if PB $>$ PCo or Net Perceived Benefit (NPB) is positive for $0<SD<PSD$:

\[
NPB = PB - PCo > 0
\]

Substitute  $PB = f(+PUFB,+PPB,+PFBP)$

\[
NPB = f(+PUFB,+PPB,+PFBP) - PCo
\]

Substitute $PPB = f(+PIPPB, +CIMEPPB)$

\[
NPB = f(+PUFB,+PIPPB, +CIMEPPB,+PFBP) - PCo
\]

Substitute $PIPPB = f(+PI)$

\[
NPB = f(+PUFB,+PI, +CIMEPPB,+PFBP) - PCo
\]

Substitute $CIMEPPB = f(+CI, +SD)$

\[
NPB = f(+PUFB,+PI, +CI, +SD,+PFBP) - PCo
\]

Substitute  $PI = f(+CI, +Epa, +Epf)$

\[
NPB = f(+PUFB, +Epa, +Epf, +CI, +SD,+PFBP) - PCo
\]

Substitute $PCo = f(+OEI, -PT)$

\[
NPB = f(+PUFB, +Epa, +Epf, +CI, +SD,+PFBP) - f(+OEI, -PT)
\]

Substitute $OEI =f(-PEC, +CI, +SD)$

\[
NPB = f(+PUFB, +Epa, +Epf, +CI, +SD,+PFBP) - f(-PEC, +CI, +SD, -PT)
\]

\[
NPB = f(+PUFB, +Epa, +Epf, +CI, +SD,+PFBP) + f(+PEC, -CI, -SD, +PT)
\]

The independent variables for calculating NPB are PUFB, Epa, Epf, PFBP, PEC and PT.

Variables PUFB, Epa, Epf, PFBP, PEC and PT exert a positive effect on NPB whereas variables CI and SD exhibit a bidirectional influence on NPB, such that CI, SD can either increase or decrease NPB.

\section*{Inter-session decision making model}

Figure 4 illustrates a computational model of inter-session decision-making. Session participation occurs if Perceived Future Cost (PFC) exceeds Perceived Future Benefit (PFB).

\begin{figure}[h!]
        \centering
        \includegraphics[width=0.8\textwidth]{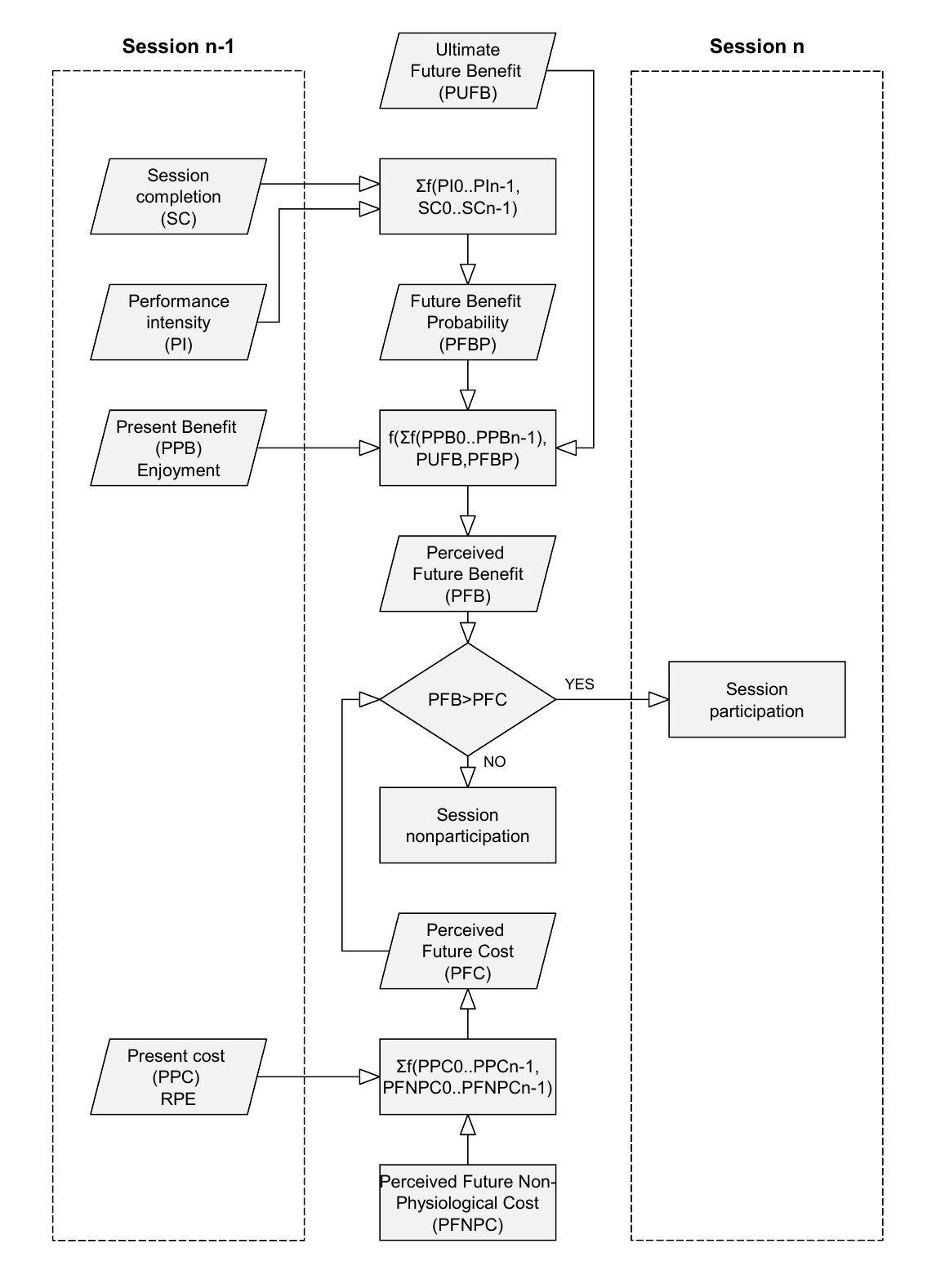}
        \caption{Inter-session decision-making.}
        \label{Figure 4}
    \end{figure}
\vspace{1em}

Inter-session Perceived Future Cost (PFC) is modeled as a progressively weighted function of all prior session Perceived Present Cost (PPC or RPE) and Perceived Future Non-Physiological Cost (PFNPC). Perceived future cost increases with an increase in prior session perceived present cost (RPE) and perceived future non-physiological cost:

\[
PFC = \sum f(PPC_{0..n-1}, PFNPC)
\]

Inter-session Perceived Future Benefit (PFB) is modeled as a progressively weighted function of all prior session Perceived Present Benefit (PPB), Perceived Ultimate Future Benefit (PUFB) and Perceived Future Benefit Probability (PFBP). Perceived future benefit increases with an increase in prior session perceived present benefit, perceived ultimate future benefit and perceived future benefit probability:

\[
PFB = f\left(\sum f(PPB_{0..n-1}), PUFB, PFBP\right)
\]

Perceived Future Benefit Probability (PFBP) is modeled as a progressively weighted function of all prior session Performance Intensities (PI) and all prior Session Completions (SC). Perceived future benefit probability increases with an increase in performance intensity and session completion:

\[
PFBP = \sum f(PI_{0..n-1}, SC_{0..n-1})
\]

\subsection*{Inter-session physiological adaptation}

Physiological adaptation $\Delta PEC$ is modeled as a function of initial Physiological Endurance Capacity (PECo), Objective Experienced Intensity (OEI) and Session Duration (SD):

\[
\Delta PEC = g(PECo, OEI, SD)
\]

\[
PCo = f(OEI, PT)
\]

Assuming $f$ is invertible:

\[
OEI = f(PCo, PT)
\]

Substitute $OEI = f(PCo, PT)$

\[
\Delta PEC = g(PECo, PCo, PT, SD)
\]

Substitute $PCo = RPE$

\[
\Delta PEC = g(PECo, RPE, PT, SD)
\]
\section*{Theoretical physical activity training intervention design for adherence}

When selecting an intensity measure for prescribing for adherence, we have four options to consider — CI, PI, OEI and subjective experienced intensity (RPE).

\bigskip

Figure 5 illustrates two scenarios, exercise prescription at constant $CI = k$ and at maximal constant $RPE = RPE_{max}$. We choose optimal $RPE_{max}$ as the maximum RPE that can be sustained over the session without resulting in session abortion. This limit is determined by the value of the perceived benefit (PB).

\begin{figure}[h!]
        \centering
        \includegraphics[width=0.8\textwidth]{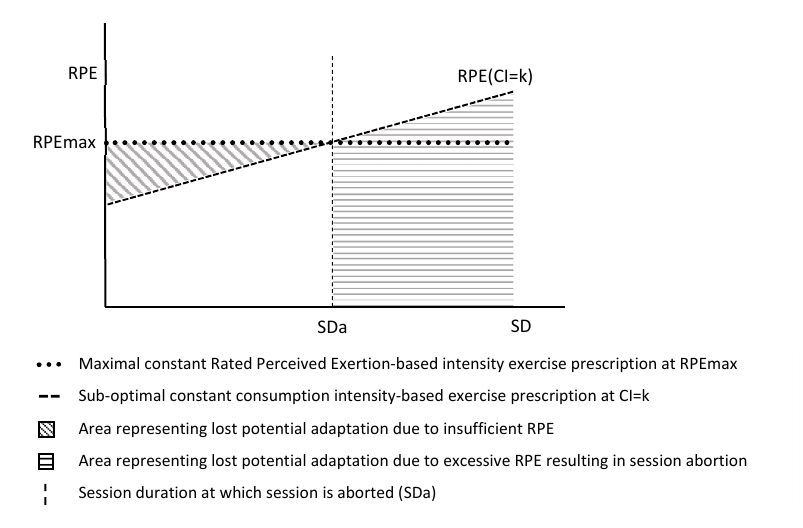}
        \caption{Conceptual representation of constant RPE versus constant CI exercise prescription.}
        \label{Figure 5}
    \end{figure}

Prescribing constant CI intensity results in RPE increasing linearly over the session duration.\textsuperscript{22} This has the consequence of RPE being sub-optimal during the first stage of the session before increasing beyond optimal RPE in the latter stage of the session, resulting in premature session abortion. Maximal physiological adaptation $\Delta PEC_{max}$ occurs when the area under the function $g(PEC, RPE, PT, SD)$ is maximised for a given completed session duration $SD$. The area under $RPE_{max}$ exceeds that of the area under $RPE(CI = k)$ for the completed session duration $SD$. Prescribing exercise intensity based on constant RPE results in greater physiological adaptation than prescribing it based on constant CI. Performance Intensity is directly proportional to CI and its prescription similarly results in sub-optimal physiological adaptation. OEI cannot practically be prescribed and additionally does not account for individual pain tolerance, resulting in inferior utility in prescription for adherence. Consequently constant RPE is selected as the optimum prescription intensity method.

\begin{figure}[h!]
        \centering
        \includegraphics[width=0.85\textwidth]{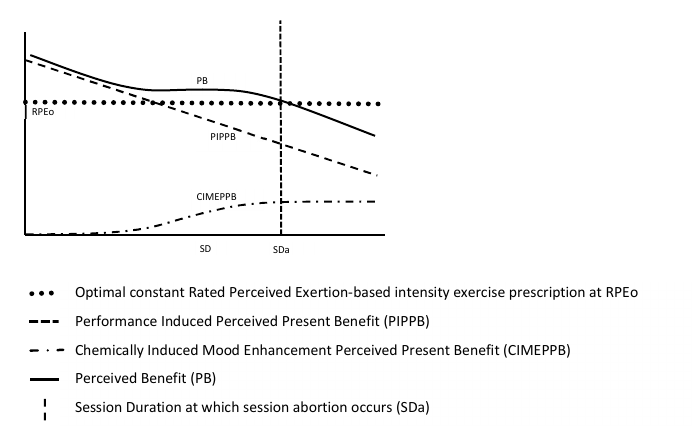}
        \caption{Conceptual session abortion at Perceived Cost (RPE) and Perceived Benefit (PB) intersection.}
        \label{Figure 6}
    \end{figure}

 Perceived Benefit is expressed as a function of PUFB, PIPPB, CIMEPPB, PFBP. 

\[
PB = f(PUFB, PPB, PFBP) = f(PUFB, PIPPB, CIMEPPB, PFBP)
\]

Only performance-induced perceived present benefit PIPPB and mood enhanced perceived present benefit CIMEPPB are functions of SD. Fig 6 illustrates a conceptual diagram of RPE, PIPPB, CIMEPPB and PB over session duration SD. For illustration purposes the effect of perceived ultimate future benefit and perceived future benefit probability is ignored. As previously discussed consumption intensity decreases linearly over session duration SD for constant RPE. Performance intensity is proportional to consumption intensity for constant exercise economy. For purposes of illustration it is assumed that performance induced perceived present benefit PIPPB will decrease proportionally with performance intensity. It is established that the release of mood-enhancing neurochemicals during physical exercise exhibits a relationship that approximates the curve characteristic of Chemically Induced Mood-Enhanced Perceived Present Benefit (CIMEPPB).\textsuperscript{23--25} For the purposes of conceptual modeling, Perceived benefit (PB) is presented as the sum of CIMEPPB and PIPPB. Assuming that PB is a decreasing function, session abortion will occur where PB and RPE intersect at session duration SDa. The function $PB = f(SD)$ acts as a constraint function for the maximum value of RPE over session duration SD:

\[
RPE \le f(SD)
\]

Utilizing the intra-session decision-making model illustrated in Fig.~3 the function $f(SD)$ is obtained as follows:

Session abortion at session duration $= SD_a$ will occur where $PCo(SD) = PB(SD)$

\[
PCo = PB
\]

Substitute $PB = f(PUFB, PPB, PFBP)$

\[
PCo = f(PUFB, PPB, PFBP)
\]

Substitute $PPB = f(PIPPB, CIMEPPB)$

\[
PCo = f(PUFB, PIPPB, CIMEPPB, PFBP)
\]

Substitute $PIPPB = f(PI)$

\[
PCo = f(PUFB, PI, CIMEPPB, PFBP)
\]

Substitute $CIMEPPB = f(CI, SD)$

\[
PCo = f(PUFB, PI, CI, SD, PFBP)
\]

Substitute $PI = f(CI, Epa, Epf)$

\[
PCo = f(PUFB, Epa, Epf, CI, SD, PFBP)
\]

\[
OEI = f(PEC, CI, SD)
\]

Assuming $f$ is invertible:

\[
CI = f(PEC, OEI, SD)
\]

Substitute $CI = f(PEC, OEI, SD)$

\[
PCo = f(PUFB, Epa, Epf, PEC, OEI, SD, PFBP)
\]

\[
PCo = f(OEI, PT)
\]

Assuming $f$ is invertible:

\[
OEI = f(PCo, PT)
\]

Substitute $OEI = f(PCo, PT)$

\[
PCo = f(PUFB, Epa, Epf, PEC, PCo, PT, SD, PFBP)
\]

Solve for $PCo$

\[
PCo = f(PUFB, Epa, Epf, PEC, PT, PFBP, SD)
\]

\[
RPE = PCo = f(PUFB, Epa, Epf, PEC, PT, PFBP, SD) = f(SD)
\]

This function represents all the values of RPE and SD for which RPE is equal to PB, for given values of PUFB, Epa, Epf, PEC, PT, PFBP.

For adherence PB must equal or exceed RPE for the duration of the session:

\[
RPE \le f(SD) = f(PUFB, Epa, Epf, PEC, PT, PFBP, SD)
\]

The objective of the training intervention is to maximize physiological adaptation while ensuring sustained adherence to the intervention.

\begin{figure}[h!]
        \centering
        \includegraphics[width=1.0\textwidth]{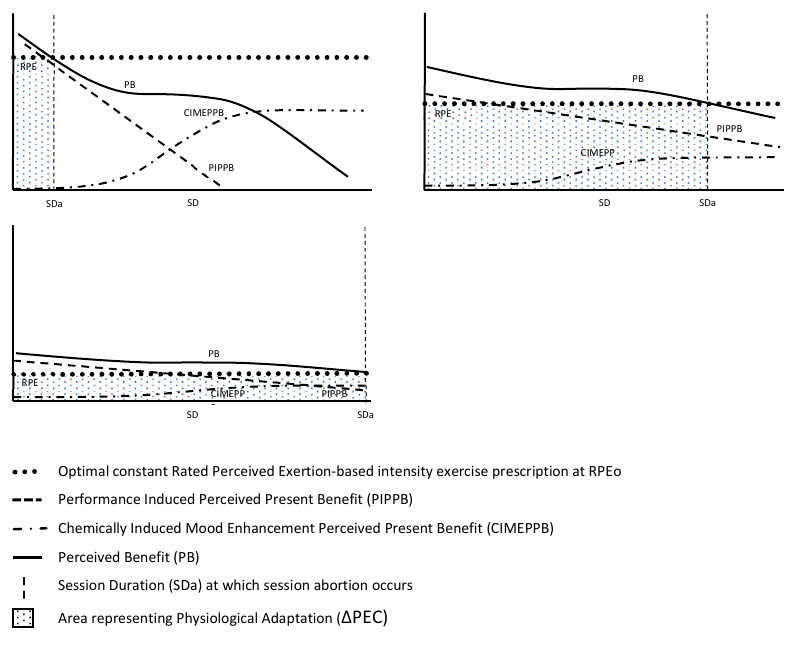}
        \caption{Conceptual representation of changes in physiological adaptation for different values of RPE.}
        \label{Figure 7}
    \end{figure}

In Figure 7 physiological adaptation, denoted by $\Delta PEC$, is represented by the shaded area
bounded by $RPE$ and $SD = SD_a$. For optimal values of perceived exertion $RPE = RPE^{*}$
and session duration $SD = SD^{*}$, physiological adaptation $\Delta PEC$ is maximized.

Taking into account the adherence constraint:
\begin{equation}
    RPE \leq f(SD),
    \label{eq:constraint}
\end{equation}
the objective function is:
\begin{equation}
    \Delta PEC = g(PEC, RPE, PT, SD) = g(RPE, SD),
    \label{eq:objective}
\end{equation}
which must be maximized subject to the constraint $RPE \leq f(SD)$.

Assuming that $PEC$, $PT$, $PUFB$, $E_{pa}$, $E_{pf}$, and $PFBP$ are known, we solve for optimal
values $RPE^{*}$ and $SD^{*}$ that maximize $\Delta PEC$ by defining the Lagrangian:
\begin{equation}
    \mathcal{L}(RPE, SD, \lambda) = g(RPE, SD) + \lambda \cdot (f(SD) - RPE).
    \label{eq:lagrangian}
\end{equation}

The necessary conditions for optimality are:
\begin{align}
    \frac{\partial \mathcal{L}}{\partial RPE} &= \frac{\partial g}{\partial RPE} - \lambda = 0, \label{eq:opt1} \\
    \frac{\partial \mathcal{L}}{\partial SD}  &= \frac{\partial g}{\partial SD} + \lambda \cdot f'(SD) = 0, \label{eq:opt2} \\
    RPE - f(SD) &\leq 0, \label{eq:opt3} \\
    \lambda &\geq 0, \label{eq:opt4} \\
    \lambda \cdot (f(SD) - RPE) &= 0. \label{eq:opt5}
\end{align}

Solving the above yields optimal values $RPE^{*}$ and $SD^{*}$ satisfying:
\begin{align}
    \frac{\partial g(RPE, SD)}{\partial RPE} \bigg|_{RPE^{*}, SD^{*}} + \lambda &= 0, \label{eq:sol1} \\
    \frac{\partial g(RPE, SD)}{\partial SD} \bigg|_{RPE^{*}, SD^{*}} - \lambda \cdot f'(SD^{*}) &= 0, \label{eq:sol2} \\
    \lambda &\geq 0, \label{eq:sol3} \\
    \lambda \cdot (RPE^{*} - f(SD^{*})) &= 0. \label{eq:sol4}
\end{align}

The optimal values $RPE^{*}$ and $SD^{*}$ define the session perceived exertion and session duration, respectively, that maximize physiological adaptation $\Delta PEC$ without risk of session abortion prior to completion of the prescribed session duration.

These values are personalized for each session by using the equations:
\begin{align}
    RPE &\leq f(SD) = f(PUFB, E_{pa}, E_{pf}, PEC, PT, PFBP, SD), \label{eq:impl1} \\
    \Delta PEC &= g(PEC, RPE, PT, SD), \label{eq:impl2}
\end{align}
resulting in an individualized exercise prescription that accounts for the following factors:

\begin{itemize}
    \item \textbf{Perceived Ultimate Future Benefit (PUFB):} Perceived benefit of completing the training intervention.
    \item \textbf{Perceived Future Benefit Probability (PFBP):} A measure of self-efficacy.
    \item \textbf{Physical Activity Efficiency ($E_{pa}$):} A constant representing physiological effort efficiency.
    \item \textbf{Performance Efficiency ($E_{pf}$):} Represents biomechanical efficiency.
    \item \textbf{Pain Tolerance ($PT$):} Individual threshold for tolerating exercise discomfort.
    \item \textbf{Physiological Endurance Capacity ($PEC$):} A conceptually equivalent measure to overall conditioning or fitness.
\end{itemize}

\begin{figure}[h!]
        \centering
        \includegraphics[width=1.02\textwidth]{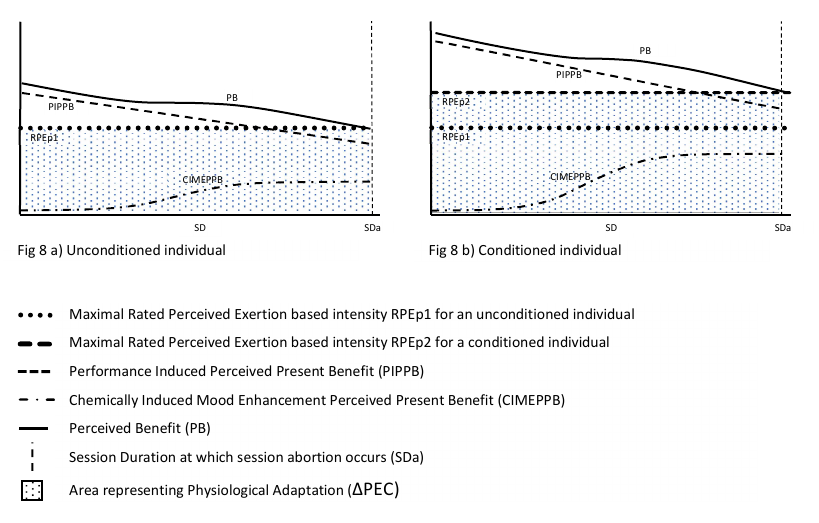}
        \caption{Conceptual physiological adaptation for different levels of conditioning (endurance capacity).}
        \label{Figure 8}
    \end{figure}

It is well established that a conditioned individual can sustain higher consumption and consequently performance intensities for a given RPE than unconditioned individuals. Neuro-chemically induced perceived benefit increases with consumption intensity.\textsuperscript{18--21} For a given RPE an unconditioned individual will perceive lower exercise related performance and mood benefits than a conditioned (higher physiological endurance capacity) individual. Our model predicts that a conditioned individual will support a higher RPE without session abortion for a given session duration than an unconditioned individual. For optimal exercise prescription, RPE prescription for an individual must be increased as physiological endurance capacity increases (PEC). This required increase is accounted for in the constraint function:

\[
RPE \le f(SD) = f(PUFB, Epa, Epf, PEC, PT, PFBP, SD)
\]

$f(SD)$ is a function of physiological endurance capacity (PEC) which increases over time between sessions of a training intervention.

Physiological endurance capacity is calculated recursively for each session:

\[
PEC_n = f(PEC_{n-1}, OEI_{n-1}, SD_{n-1}) = \sum f(PEC_0..PEC_{n-1}, OEI_0..OEI_{n-1}, SD_0..SD_{n-1})
\]

Where $OEI_{n-1} = f(RPE^*_{n-1}, PT)$ and $SD_{n-1} = SD^*_{n-1}$

$RPE^*_n$ and $SD^*_n$ are calculated recursively for each session from $PEC_n$.

It is well established that physiological adaptation $\Delta PEC$ decreases exponentially with conditioning for a given objective experienced intensity and session duration.

Figure 9 conceptually illustrates declining session physiological adaptation $\Delta PEC$ for an increase in physiological endurance capacity $PEC_n$.\textsuperscript{27} The total number of sessions for the training intervention will be limited to some value of $n =$ Total Session Count (TSC) where $\Delta PEC_n / \Delta PEC_0 \le k$. Training intervention total duration (TD) in weeks is calculated by setting $k$ and calculating TSC recursively. TSC is divided by weekly session frequency.

\begin{figure}[h!]
        \centering
        \includegraphics[width=0.8\textwidth]{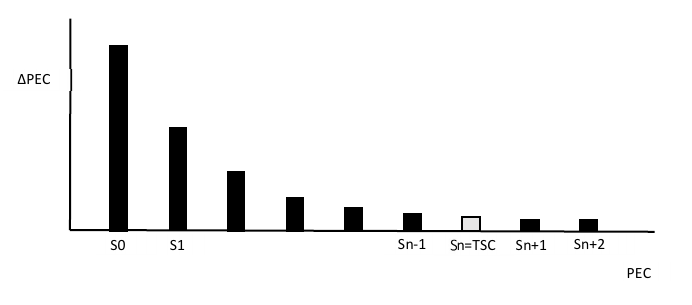}
        \caption{Conceptual session physiological adaptation for different levels of conditioning (PEC).}
        \label{Figure 9}
    \end{figure}

\section*{Practical physical activity training intervention design for adherence}

To enable the practical design of a training intervention for an individual, it is necessary to determine the values of relevant variables. Theoretically defined functions must be developed into equations with explicit, operational forms.\\

Values of training intervention variables that require determination:\\
\\Perceived ultimate future benefit PUFB (perceived benefit of completing the training intervention)
\\Physical activity efficiency Epa (assumed to be an inter and intra-individual constant k)\textsuperscript{26}
\\Performance energy efficiency Epf (biomechanical efficiency)
\\Pain tolerance PT
\\Physiological endurance capacity PEC (conceptually equivalent to “conditioning” or “fitness”)\\

Training intervention functions:

\[
\begin{aligned}
PI &= f(CI, Epa, Epf) \\
PEC_n &= f(PEC_{n-1}, OEI_{n-1}, SD_{n-1}) \\
\Delta PEC &= g(PEC_0, OEI, SD) \\
PCo &= RPE = f(OEI, PT) \\
OEI &= f(PEC, CI, SD) \\
PB &= f(PUFB, PPB, PFBP) \\
PFBP &= \Sigma f(n, PI_0..PI_{n-1}, SC_0..SC_{n-1}) \\
PPB &= f(PIPPB, CIMEPPB) \\
PIPPB &= f(PI) \\
CIMEPPB &= f(CI, SD)
\end{aligned}
\]

\subsection*{Determination of training intervention design variables}

Perceived Ultimate Future Benefit (PUFB) can be assessed through the use of a questionnaire. The authors intend to conduct further research aimed at developing and validating such an instrument.

Physical activity efficiency (Epa) has been measured and estimated using a variety of methodological approaches. However, there is no conclusive evidence to suggest that it varies as a function of conditioning.\textsuperscript{28--31} For the purposes of intervention design, it will therefore be assumed to represent an inter- and intra-individual constant, denoted as $K_{pa} \approx 30\%$.\textsuperscript{26}

Performance energy efficiency Epf (bio-mechanical efficiency) is activity specific. In a laboratory environment Epf can be calculated from CI and PI. The authors intend to conduct further research aimed at developing and validating activity specific methodologies for measuring individual bio-mechanical efficiency in practical intervention training groups.

The authors intend to conduct further research aimed at developing and validating a range of methodologies to measure and predict Pain tolerance (PT).

Physiological Endurance Capacity (PEC) can be estimated by applying the proposed recursive theoretical model ($PEC_n = f(PEC_{n-1}, OEI_{n-1}, SD_{n-1})$), which allows the calculation of endurance capacity in reverse, based on self-reported physical activity session intensity (RPE), duration (SD) and frequency.

\subsection*{Development of training intervention design equations}

\underline{Proposed methodology for regression modeling data collection}

\begin{enumerate}
\item A baseline, untrained sample population will complete established pain tolerance questionnaires (e.g., PSQ, PRS, XX).
\item The sample population will then be subjected to repeated training sessions on a cycle ergometer at a prescribed RPE intensity level (RPEx) and session duration (SDz) twice a week.
\item Before each session the participants will be asked to complete Perceived Ultimate Future Benefit (PUFB), Perceived Future Benefit Probability (PFBP), Success Perception (SP), Achievement Perception (AP) and Improvement Perception (IP) questionnaires.
\item VO$_2$, VCO$_2$, heart rate and power output will be continuously recorded during every fourth session.
\item Participants will be asked to complete Perceived Benefit (PB), Perceived Present Benefit (PPB), Performance Induced Perceived Present Benefit (PIPPB) and Chemically Induced Mood Enhancement Perceived Present Benefit (CIMEPPB) questionnaires every 5 minutes during the session for every second session.
\item The same procedure will be repeated for multiple sample populations by substituting all value permutations of RPE and SD of interest into RPEx and SDz, respectively.
\end{enumerate}

\underline{Modeling Performance Intensity (PI)}

The relationship between performance intensity (PI), consumption intensity (CI), physical activity efficiency (Epa), and physiological energy efficiency (Epf) can be expressed as follows:

\[
PI = f(CI, Epa, Epf)
\]

By definition of the component variables, this reduces to a multiplicative relationship:

\[
PI = CI \times Epa \times Epf
\]

\underline{Modeling Physiological Endurance Capacity (PEC)}

Physiological endurance capacity change ($\Delta PEC$) can be represented as a function of initial endurance capacity (PEC$_0$), objective experienced intensity (OEI), and session duration (SD):

\[
\Delta PEC = g(PEC_0, OEI, SD)
\]

At present, direct measurement of OEI is not possible. OEI is instead conceptualized as a function of subjective experienced intensity, operationalized through the Rated Perceived Exertion (RPE) scale, and moderated by pain tolerance (PT). Individual PT values acquire meaning only in relation to the statistical mean of a reference population. For sufficiently large populations, the mean OEI can therefore be approximated by the mean RPE. Accordingly, where a statistical relationship between $\Delta PEC$ and OEI is derived, OEI may be substituted with RPE:

\[
\Delta PEC = g(PEC_0, OEI, SD) = g(PEC_0, RPE, SD)
\]

The initial physiological endurance capacity (PEC$_0$) will be defined as the mean consumption energy (energy expenditure) of the sample population measured during the first session. The mean physiological endurance capacity $PEC_{(n*4)}(RPE = RPEx, SD = SDz)$ for the sample population will be calculated for every fourth session. These values will be used to construct a discrete regression equation of the form:

\[
PEC_n(RPE = RPEx, SD = SDz) = f_{x,z}(SD_n)
\]

This equation will enable interpolation of endurance capacities across all session indices n. The same procedure will be repeated for all relevant permutations of RPE and SD. From this, a matrix of endurance capacity values $PEC_n(RPE=RPE(x_0..x), SD=SD(z_0..z))$ will be generated. This matrix will then be synthesized into a general regression equation:

\[
\Delta PEC = g(PEC_0, RPE, SD)
\]

where $\Delta PEC_n = PEC_n - PEC_{n-1}$.\\

\underline{Modeling Subjective Experienced Intensity (RPE)}

Physiological adaptation in individuals surpassing the population mean for a given rate of perceived exertion (RPE) and energy expenditure over the session duration (SD), may be attributed to either genetic predisposition or an elevated pain tolerance. An increased pain tolerance is associated with a higher objective experienced intensity for a given RPE, leading to greater physiological adaptation. The authors propose that cohorts within a sample population demonstrating superior physiological adaptation for a particular RPE will contain a larger proportion of individuals with higher pain tolerance, while the reverse will be true for cohorts exhibiting lower levels of physiological adaptation. Furthermore, the authors hypothesize that heart rate (HR) and blood lactate (La) can serve as objective indicators of objective experienced intensity. Correlations between physiological adaptation, HR, and La will be cross-validated with results from established pain tolerance questionnaires (e.g., PSQ, PRS). This data will be further analyzed to assess relationships between pain tolerance scores and HR and La levels within subgroups of the respective cohorts. A strong correlation between pain tolerance scores from questionnaires and objective physiological markers such as HR and La would support the conclusion that the greater physiological adaptation observed in certain individuals is a result of higher pain tolerance rather than genetic factors. The authors aim to develop a specific exercise-related pain tolerance questionnaire that could serve as a reliable predictor of pain tolerance (PT) within the context of their training intervention model. Successful implementation of this methodology would yield a dataset encompassing objective experienced intensity (OEI), RPE and pain tolerance (PT), which could be used to construct a regression equation of the form:

\[
PCo = RPE = f(OEI, PT)
\]

\underline{Modeling Objective Experienced Intensity (OEI)}

The function:

\[
OEI = f(PEC, CI, SD)
\]

is employed within the training intervention model with the objective of deriving its inverse form:

\[
CI = f(PEC, OEI, SD)
\]

As previously established, for sufficiently large sample populations, the mean Objective Experienced Intensity (OEI) can be approximated by the mean Rated Perceived Exertion (RPE). Consequently, when a statistical relationship between Consumption Intensity (CI) and OEI is determined, OEI may be substituted with RPE:

\[
CI = f(PEC, OEI, SD) = f(PEC, RPE, SD)
\]

Discrete values of CI, denoted as $CI = CI_k$, will be computed at one-minute intervals of the session duration, represented as $SD = SD_k$. For each given value of RPE (RPEx), physiological endurance capacity ($PEC = PEC_{n,x}$), and session duration interval ($SD = SD_k$), the mean consumption intensity for the population will be calculated. This process will be iterated across all combinations of x and n, corresponding to RPE levels and training session counts, respectively.

From these calculations, a matrix of consumption intensity values will be constructed in the form:

\[
CI(PEC = PEC((x_0,n_0) \ldots (x,n)), RPE = RPE(x_0 \ldots x), SD = SD(k_0 \ldots k))
\]

This matrix will subsequently be used to derive a generalized regression equation:

\[
CI = f(PEC, RPE, SD)
\]

\underline{Modeling Perceived Benefit (PB)}

Experimental data collected through the Perceived Benefit (PB), Perceived Ultimate Future Benefit (PUFB), Perceived Present Benefit (PPB), and Perceived Future Benefit Probability (PFBP) questionnaires will be normalized to a standardized scale ranging from 1 to 20. The resulting data matrix will then be utilized to derive a generalized regression equation:

\[
PB = f(PUFB, PPB, PFBP)
\]

\underline{Modeling Perceived Future Benefit Probability (PFBP)}

The proposed training intervention model conceptualizes Perceived Future Benefit Probability (PFBP) as follows:

\[
PFBP = \Sigma f(n, PI_0..PI_{n-1}, SC_0..SC_{n-1})
\]

Within this framework, the authors posit that PFBP—functionally analogous to the construct of self-efficacy—is a complex function derived from individuals’ perceptions of their past performance. Specifically, the model identifies the following primary components contributing to PFBP:

\begin{enumerate}
\item Success Perception (SP): Defined as a function of Session Completion.
\item Achievement Perception (AP): Defined as a function of Session Performance.
\item Improvement Perception (IP): Defined as a function of the change in Session Performance over time.
\end{enumerate}

Further empirical investigation is necessary to determine the optimal computational methodology for modeling the PFBP function. To this end, data will be collected through targeted questionnaires designed to assess SP, AP, and IP. The resulting data will inform the development of a formal computational model of Perceived Future Benefit Probability.\\

\underline{Modeling Perceived Present Benefit (PPB)}

Experimental data collected from the Perceived Present Benefit (PPB), Performance-Induced Perceived Present Benefit (PIPPB), and Chemically Induced Mood Enhancement Perceived Present Benefit (CIMEPPB) questionnaires will be normalized to a standardized scale ranging from 1 to 20. The resulting data matrix will then be utilized to derive a generalized regression equation:

\[
PPB = f(PIPPB, CIMEPPB)
\]

\underline{Modeling Performance Induced Perceived Present Benefit (PIPPB)}

Experimental data collected from the Performance Induced Perceived Present Benefit (PIPPB) questionnaires will be normalized to a standardized scale ranging from 1 to 20. Corresponding measurements of Performance Intensity (PI), aligned with the time intervals of questionnaire administration, will be identified. The resulting data matrix will then be utilized to develop a generalized regression equation:

\[
PIPPB = f(PI)
\]

\underline{Modeling Chemically Induced Mood Enhancement Perceived Present Benefit (CIMEPPB)}

Experimental data collected through the Chemically Induced Mood Enhancement Perceived Present Benefit (CIMEPPB) questionnaires will be normalized to a standardized scale ranging from 1 to 20. Corresponding measurements of Consumption Intensity (CI), recorded at time points coinciding with the questionnaire administration intervals $SD = SD_k$, will be identified. Based on these data, a matrix comprising the normalized CIMEPPB values will be constructed in the following form:

\[
CIMEPPB(CI=CI(k_0 \ldots k), SD=SD(k_0 \ldots k))
\]

This data matrix will subsequently be used to derive a generalized regression equation:

\[
CIMEPPB = f(CI, SD)
\]

\subsection*{Conclusion}

This paper highlights critical conceptual and methodological limitations in the current understanding and application of VO$_2$max as a measure of cardiorespiratory function. Specifically, VO$_2$max reflects cardiorespiratory capacity rather than cardiorespiratory endurance, and even if it did represent cardiorespiratory endurance, it remains a poor predictor of actual endurance performance. Furthermore, the lack of standardized definitions for VO$_2$max and metabolic equivalents (METs) relative to lean body mass within existing ACSM guidelines introduces significant inaccuracies in both epidemiological research and individualized physical activity assessment and prescription. The utility of VO$_2$max as a clinical diagnostic tool is also called into question, given that its ability to predict long term risk of mortality incurs concomitant short-term risk of mortality associated with its administration. Current definitions of physical activity intensity and fitness are insufficiently precise, confounded by multiple variables, and thus inadequate for advancing health science in the domains of physical activity measurement and prescription. To address these issues, the authors propose revised, operationally precise definitions of physical activity fitness and intensity. These refined definitions underpin the development of a rigorous mathematical and computational model for training prescription. Successful validation of this model holds significant promise for leveraging computational tools to enhance both adherence to and efficacy of health-oriented training programs.

\section*{References}

1. Liguori G. ACSM's Guidelines for Exercise Testing and Prescription. 11th ed. Philadelphia, PA: Wolters Kluwer Health; 2022.\\
2. Coyle EF, Coggan AR, Hopper MK, Walters TJ. Determinants of endurance in well-trained cyclists. J Appl Physiol (1985) 1988; 64(6): 2622-30.\\
3. Vollaard NB, Constantin-Teodosiu D, Fredriksson K, et al. Systematic analysis of adaptations in aerobic capacity and submaximal energy metabolism provides a unique insight into determinants of human aerobic performance. J Appl Physiol (1985) 2009; 106(5): 1479-86.\\
4. Clarke J, de Lannoy L, Ross R. Comparison of Measures of Maximal and Submaximal Fitness in Response to Exercise. Med Sci Sports Exerc 2017; 49(4): 711-6.\\
5. Millet GP, Vleck VE, Bentley DJ. Physiological differences between cycling and running: lessons from triathletes. Sports Med 2009; 39(3): 179-206.\\
6. Ainsworth BE, Haskell WL, Herrmann SD, et al. 2011 Compendium of Physical Activities: a second update of codes and MET values. Med Sci Sports Exerc 2011; 43(8): 1575-81.\\
7. Herrmann SD, Willis EA, Ainsworth BE, et al. 2024 Adult Compendium of Physical Activities: A third update of the energy costs of human activities. J Sport Health Sci 2024; 13(1): 6-12.\\
8. Borrud LG, Flegal KM, Looker AC, Everhart JE, Harris TB, Shepherd JA. Body composition data for individuals 8 years of age and older: U.S. population, 1999-2004. Vital Health Stat 11 2010; (250): 1-87.\\
9. Williams PT. Physical fitness and activity as separate heart disease risk factors: a meta-analysis. Med Sci Sports Exerc 2001; 33(5): 754-61.\\
10. Global BMIMC, Di Angelantonio E, Bhupathiraju Sh N, et al. Body-mass index and all-cause mortality: individual-participant-data meta-analysis of 239 prospective studies in four continents. Lancet 2016; 388(10046): 776-86.\\
11. Boning DM, N.; Steinach, M. The efficiency of muscular exercise. Deursche Zeitschrift Fur Sportmedizin 2017; 68(9): 203-13.\\
12. Warburton DE, Charlesworth S, Ivey A, Nettlefold L, Bredin SS. A systematic review of the evidence for Canada's Physical Activity Guidelines for Adults. Int J Behav Nutr Phys Act 2010; 7: 39.\\
13. Wen CP, Wai JP, Tsai MK, et al. Minimum amount of physical activity for reduced mortality and extended life expectancy: a prospective cohort study. Lancet 2011; 378(9798): 1244-53.\\
14. Matei D, Trofin D, Iordan DA, et al. The Endocannabinoid System and Physical Exercise. Int J Mol Sci 2023; 24(3).\\
15. Heijnen S, Hommel B, Kibele A, Colzato LS. Neuromodulation of Aerobic Exercise-A Review. Front Psychol 2015; 6: 1890.\\
16. Dietrich A, McDaniel WF. Endocannabinoids and exercise. British Journal of Sports Medicine 2004; 38(5): 536-41.\\
17. Pilozzi A, Carro C, Huang X. Roles of beta-Endorphin in Stress, Behavior, Neuroinflammation, and Brain Energy Metabolism. Int J Mol Sci 2020; 22(1).\\
18. Rojas Vega S, Strüder HK, Vera Wahrmann B, Schmidt A, Bloch W, Hollmann W. Acute BDNF and cortisol response to low intensity exercise and following ramp incremental exercise to exhaustion in humans. Brain Research 2006; 1121(1): 59-65.\\
19. Siebers M, Biedermann SV, Fuss J. Do Endocannabinoids Cause the Runner's High? Evidence and Open Questions. Neuroscientist 2023; 29(3): 352-69.\\
20. Ferris LT, Williams JS, Shen CL. The effect of acute exercise on serum brain-derived neurotrophic factor levels and cognitive function. Med Sci Sports Exerc 2007; 39(4): 728-34.\\
21. Goldfarb AHH, B. Plasma beta-endorphin concentration: Response to intensity and duration of exercise. Medicine \& Science in Sports \& Exercise 1990; 22(2): 241-4.\\
22. Monteiro W, Cunha F, Brasil I, Joi S, Farinatti P. Rates of Perceived Exertion Obtained From Cardiopulmonary Exercise Testing Are Not Reproduced during Prolonged Aerobic Bouts. Journal of Exercise Physiology 2019; 22(4): 29-38.\\
23. Goldfarb AHH, B.D.; Potts, J.; Armstrong, D. Beta-Endorphin Time Course Response to Intensity of Exercise: Effect of Training Status. International Journal of Sports Medicine 1991; 12(3): 264-8.\\
24. Rasmussen P, Brassard P, Adser H, et al. Evidence for a release of brain-derived neurotrophic factor from the brain during exercise. Exp Physiol 2009; 94(10): 1062-9.\\
25. Sparling PB, Giuffrida A, Piomelli D, Rosskopf L, Dietrich A. Exercise activates the endocannabinoid system. Cognitive Neuroscience and Neuropsychology 2003; 14(17): 2209-11.\\
26. Macintosh BR. Chapter 6A: Energy Supply for Exercise In: Macintosh BR, editor. Open Textbook of Exercise Physiology. Calgary, Alberta: University of Calgary.\\
27. Hellsten Y, Nyberg M. Cardiovascular Adaptations to Exercise Training. Comprehensive Physiology 2016; 6: 1-32.\\
28. Böning D, Maassen N, Steinach M. The efficiency of muscular exercise. Deutsche Zeitschrift für Sportmedizin 2017; 2017(09): 203-14.\\
29. Montero D, Cathomen A, Jacobs RA, et al. Haematological rather than skeletal muscle adaptations contribute to the increase in peak oxygen uptake induced by moderate endurance training. The Journal of Physiology 2015; 593(20): 4677-88.\\
30. Meinild Lundby AK, Jacobs RA, Gehrig S, et al. Exercise training increases skeletal muscle mitochondrial volume density by enlargement of existing mitochondria and not de novo biogenesis. Acta Physiol (Oxf) 2018; 222(1).\\
31. Dandanell S, Meinild-Lundby AK, Andersen AB, et al. Determinants of maximal whole-body fat oxidation in elite cross-country skiers: Role of skeletal muscle mitochondria. Scand J Med Sci Sports 2018; 28(12): 2494-504.\\

\end{document}